\documentclass[
  aps,
  pre,
  a4paper,
  english,
  reprint,
  superscriptaddress,
  showkeys
]{revtex4-1}
\usepackage[T1]{fontenc}
\usepackage[utf8]{inputenc}
\usepackage{graphicx}
\usepackage[american]{babel}
\usepackage{amssymb,amsmath}
\usepackage[
  caption=false
]{subfig}
\usepackage{refstyle}
\usepackage[
  citecolor=blue,
  colorlinks,
  linkcolor=blue,
  urlcolor=blue,
]{hyperref}
\usepackage{xcolor}
\usepackage[babel=true]{csquotes}

\begin{document}

\global\long\def\pgr{\mathcal{P}_{\text{gr}}}
\global\long\def\pdb{\mathcal{P}_{\text{db}}}
\global\long\def\pov{\mathcal{P}_{\text{ov}}}
\global\long\def\pn{\mathcal{P}_{0}}
\global\long\def\df{d_{\text{f}}}

\title{Discrete Scale Invariance in Supercritical Percolation}

\author{Malte Schr\"oder}
\email{malte@nld.ds.mpg.de}
\affiliation{
  Network Dynamics,
  Max Planck Institute for Dynamics and Self-Organization (MPIDS),
  37077 G\"ottingen, Germany
}

\author{Wei Chen}
\email{w.chen@neu.edu}
\affiliation{Center for Complex Network Research, Department of Physics, Northeastern University, MA 02115, Boston, USA. }

\author{Jan Nagler}
\email{jnagler@ethz.ch}
\affiliation{
Computational Physics, IfB,
ETH Zurich, Wolfgang-Pauli-Strasse 27,
8093 Zurich, Switzerland
}

\begin{abstract}
Recently it has been demonstrated that the connectivity transition from microscopic connectivity to macroscopic connectedness, known as percolation, is generically announced by a cascade of microtransitions of the percolation order parameter [
Chen et al., Phys. Rev. Lett. 112, 155701 (2014)].
Here we report the discovery of {\em macro}transition cascades 
which follow percolation. The order parameter grows in discrete macroscopic steps with positions that can be randomly distributed even in the thermodynamic limit.
These transition positions are, however, correlated and follow scaling laws which arise from discrete scale invariance and non self-averaging, both traditionally unrelated to percolation. We reveal the discrete scale invariance in ensemble measurements of these non self-averaging systems by rescaling of the individual realizations before averaging.
\end{abstract}

\global\long\def\pgr{\mathcal{P}_{\text{gr}}}
\global\long\def\pdb{\mathcal{P}_{\text{db}}}
\global\long\def\pov{\mathcal{P}_{\text{ov}}}
\global\long\def\pn{\mathcal{P}_{0}}
\global\long\def\df{d_{\text{f}}}
\global\long\def\DCmax{G_1}
\global\long\def\Rv{\mathcal{R}_v}

\newcommand{\ie}{\emph{i.e.}}
\newcommand{\eg}{\emph{e.g.}}
\newcommand{\ER}{Erd\H{o}s-R\'{e}nyi}
\newcommand{\AP}{Achlioptas prcoesses}
\newcommand{\X}{\mathcal{O}}

\newcommand{\red}[1]{{\color{red}#1}}
\newcommand{\blue}[1]{{\color{blue}#1}}
\newcommand{\orange}[1]{{\color{orange}#1}}
\newcommand{\malte}[1]{{\color{violet}\textsc{MALTE:} \textbf{#1}}}
\newcommand{\jan}[1]{{\color{green}\textsc{JAN:} \textbf{#1}}}


\maketitle

Percolation penetrates many natural, technological and social systems \cite{StaufferPercBook, Sornette1990, DrosselPRL1992,ParshaniandBulyrevandStanley2010,CallawayPRL2000,NewmanandWatts2002, saberi13, nagler2015}, ranging from conductivity of composite materials~\cite{Sahimi,AndradePRE2000} and polymerization~\cite{ZiffPRL1982} to epidemic spreading~\cite{Anderson1991,CmoorePRE2000,PastorPRL2001} and information diffusion~\cite{StrangARS1998,Lazarsfeld1944}.
Across all percolation systems, once the density of links in the networked system exceeds a critical threshold at $p>p_c$, the system undergoes a sudden transition from microscopic connectedness to global connectivity.

This critical transition to global connectedness is typically accompanied by continuous scale invariance \cite{stanley71, sornettebook}, this means power-law scaling of observables close to the transition point \cite{StaufferPercBook}. Discrete scale invariance (DSI) arises when such continuous scale invariance is partially broken. This means the scale invariance of an observable $F(x)\sim x^{\alpha}$ obeying $\frac{F(\lambda x)}{F(x)}=\lambda^{\alpha} = \mu$ does not hold for all  $\lambda$ anymore but only for a countable set $\lambda_1,\lambda_2,...$ with a fixed $\lambda$ being the fundamental scaling ratio of the system and $\lambda_n=\lambda^n$. The discreteness of the scaling factors then leads to log-periodic modulations of the continuous scale invariance and of the power-law behavior of the observable \cite{sornetteDSI, sornettebook}.

For models of continuous and discontinuous percolation, it has recently been demonstrated that
percolation at $p=p_c$ is generically announced by a cascade of microtransitions of the size of the largest interconnected component, $S_1\rightarrow \mu S_1$ at discrete positions $p_i < p_c$, exhibiting discrete scale invariance \cite{nagler2014}.

Various percolation models exhibit non-trivial behavior also in the supercritical regime \cite{chen13unstable,chen2013phase,nagler2015}, such as multiple giant components or macroscopic, discontinuous transitions. Here we report the discovery that fractional percolation models exhibit DSI for supercritical link densities $p>p_c$. Specifically, we demonstrate DSI to be caused by the strict fractional growth mechanism and the induced delay of macroscopic scale between growth steps of the largest cluster. Fractional growth also often leads to non self-averaging in these models.

We analytically identify the relation of supercritical DSI in this class of percolation models to the standard supercritical scaling.  Due to the coexistence with non self-averaging the DSI is often hidden in standard ensemble-averaged statistics. To reveal the DSI even in these measurements we propose a rescaling method, where individual realizations are aligned before averaging (the realizations in self-averaging systems are already self-aligned). 
We numerically and, where possible, analytically exemplify these findings for different percolation models.

All models discussed here follow the same general structure: starting from an initially empty network with $N$ nodes but $L=0$ links, links are added one at a time based on the given percolation rule. We observe the size of the largest cluster $S_1$ depending on the link density in the network $p=L/N$. We specifically analyze percolation models leading to a fractional growth mechanism for the largest cluster.

We discuss two variants of homophilic percolation \cite{nagler2012, nagler2013}, introduced as the first example of fractional percolation models. Both models exhibit supercritical DSI with a simple percolation rule. First we discuss \textit{global homophilic percolation} and the supercritical DSI in this self-averaging system. Then we consider the local \textit{homophilic percolation} model, illustrating how the signs of DSI can be completely hidden in ensemble-averaged properties due to the non self-averaging, and introduce the rescaling approach. Finally, we discuss a \textit{modified \ER{}-model} \cite{riordan2012} with similar properties. Due to its relation to standard \ER{}-percolation we are able to analytically derive the DSI scaling parameters for this model. The individual models are described in more detail in their individual paragraphs.\\
\newpage
\textit{Non self-averaging in percolation}.---Non self-averaging in percolation describes the phenomenon that the evolution of the relative size of the largest cluster does not converge to a well defined function $(S_1/N)(p)$ of the link density as the system size $N \rightarrow \infty$. Instead $(S_1/N)(p)$ remains broadly distributed with a finite variance. Examples of non self-averaging percolation models were first given in \cite{riordan2012,nagler2012}. Typically, non self-averaging is defined by the relative variance of the order parameter
\begin{equation}
 \mathcal{R}_v\left[S_1(p)\right] = \frac{\left<S_1(p)^2\right> - \left<S_1(p)\right>^2}{\left<S_1(p)\right>^2} \,,
\end{equation}
where $\left<\cdot\right>$ denotes ensemble averaging. If the system is self-averaging and the relative size of the largest cluster $S_1/N$ becomes a function of $p$, then $\mathcal{R}_v\left[S_1(p)\right] \rightarrow 0$ for large systems. If, on the other hand, $\mathcal{R}_v\left[S_1(p)\right]$ does not disappear for $N \rightarrow \infty$ the system is said to be non self-averaging \cite{sornettebook}. In this case $S_1/N$ is (randomly) distributed in some interval and differences between individual realizations do not decay for large systems.\\

\textit{DSI in percolation}.---Recently DSI has been demonstrated to occur in a percolation model with global competition where in each step the two smallest clusters in the system merge \cite{nagler2014}. Start with an empty graph with $N$ isolated nodes. At each step connect the two smallest clusters in the system. If multiple choices are valid pick one uniformly at random. This model is the limiting case $m\rightarrow\infty$ of the original explosive percolation models \cite{achlioptas}, where at each step a fixed number of $m$ links compete for addition \cite{friedman2009, nagler2011}.
The global competition suppresses transitions different from {\em doubling} transitions
 $S_1\rightarrow 2 S_1$ leading to $p_c = 1$. These transitions occur after adding $L_j$ links to the network at fixed link densities $L_j/N = p_j = (2^j-1)/2^j$ and clearly announce the percolation transition as
\begin{equation}\label{GCDSI}
	p_j = p_c - 2^{-j}\,,
\end{equation}
for $j > 0$ and therefore
\begin{equation} 
	\frac{p_c-p_j}{p_c-p_{j+1}} = 2 \,, \quad\quad S_1(p_j) = \frac{1}{2} S_1(p_{j+1}) \,.
\end{equation}
These discrete jumps thus follow DSI with scaling factors $\mu = 1/2$, $\lambda = 2$ and exponent $\alpha = -1$ \cite{nagler2014}.

\begin{figure}[h]
\centering
\includegraphics[width = 0.48\textwidth]{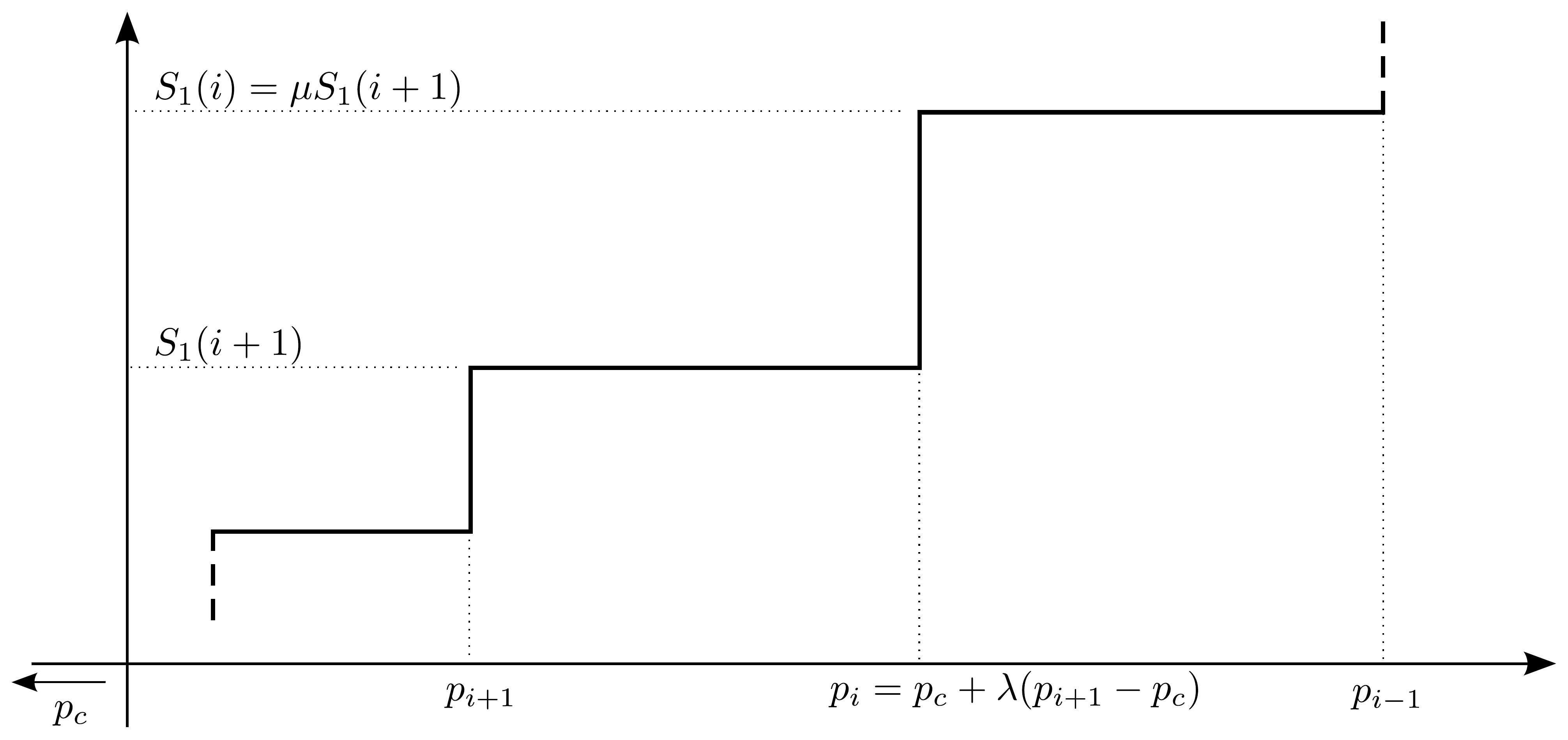}
\caption{
\label{fig:notation}
{\bf Indexing of the steps of the supercritical staircase.}
As described in the main text the steps are numbered from right to left beginning with the last (observed) step. The size of the largest cluster during these steps is indexed accordingly. With this indexing $p(i \rightarrow \infty) \rightarrow p_c$ in the thermodynamic limit and the DSI is described as in Eq.~(\ref{eq:dsi_notation}).
}
\end{figure}

Similar to this concept we discuss the \emph{supercritical} DSI using the following notation. We index the positions of the steps of $S_1$ as $p_i$ counting the steps backwards to the critical point, i.e., $p_i > p_{i+1}$. We choose this indexing since there are only a finite number of steps after any $p > p_c$ due to the obvious limit of $S_1/N \le 1$ but, in the thermodynamic limit, infinitely many jumps close to the critical point. Accordingly, we write $S_1(p_i < p < p_{i-1}) = S_1(i)$. The notation is illustrated in Fig.~\ref{fig:notation}.

With this notation the supercritical discrete scale invariance is described by 
\begin{equation}
 S_1(p_i) = S_1\left( p_c + \lambda (p_{i+1} - p_c) \right) = \mu S_1(p_{i+1})\,
\end{equation}
and therefore
\begin{equation}\label{eq:dsi_notation}
 \frac{p_i-p_c}{p_{i+1}-p_c} = \lambda \,, \quad\quad \frac{S_1(i)}{S_1(i+1)} = \mu \,\\
\end{equation}
%
For the models considered here the factor $\mu$ is given by the fractional growth mechanism of the percolation rules and $\lambda$ can be estimated using the positions of the individual steps. The DSI exponent is then given as $\alpha = \log(\mu) / \log(\lambda)$.\\

\textit{Global homophilic percolation}.---Starting from a modification of global competition we use a model with global \enquote{homophilic} competition to demonstrate that DSI can occur in the supercritical regime as well.
Again, starting from $N$ isolated nodes, at each step choose a link uniformly at random from all links joining clusters of the same size (possibly the same cluster) and add this link. We denote this model as global homophilic percolation (GHP). This modification retains the strict doubling rule (fractional growth) from the original globally competitive model but it exhibits a non-trivial supercritical regime as it does not always merge the two smallest clusters.

\begin{figure}[h]
\centering
\includegraphics[width = 0.48\textwidth]{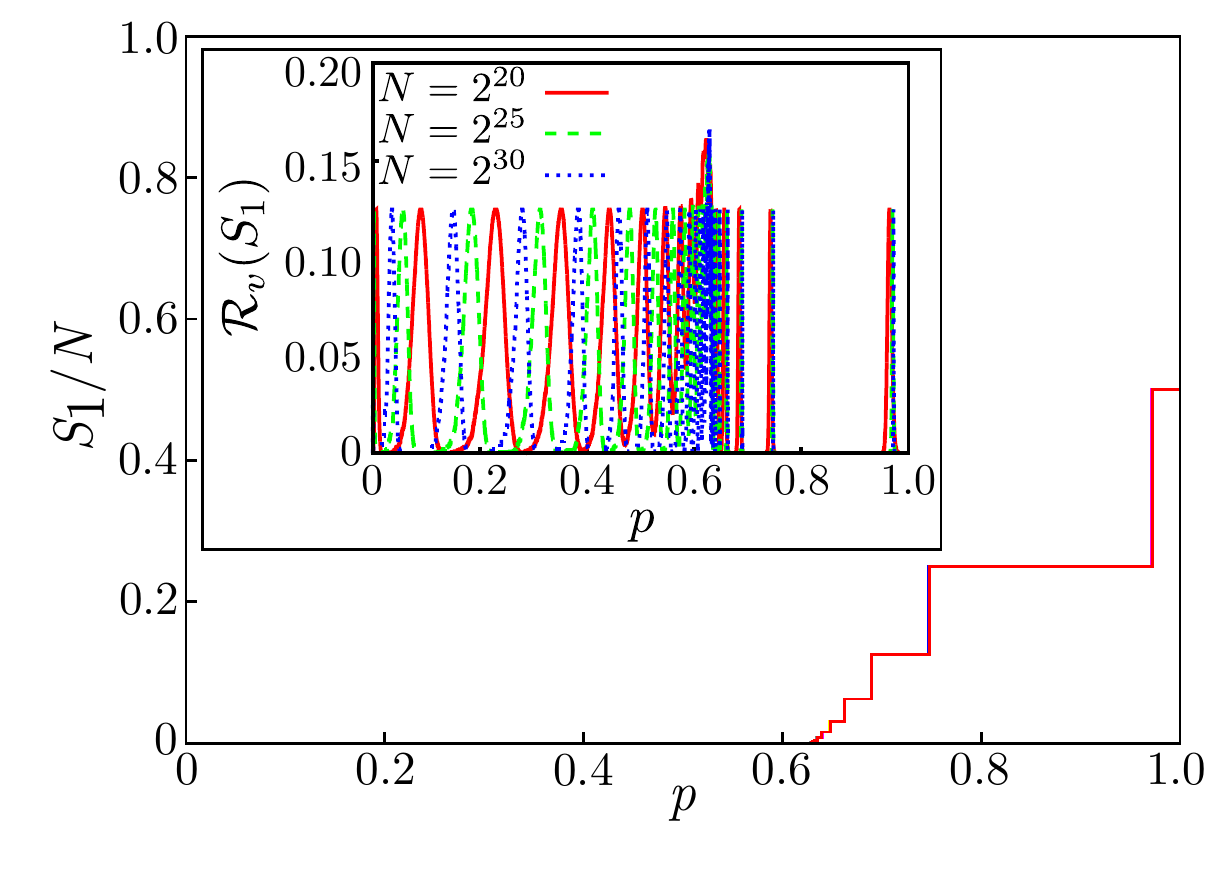}
\caption{
\label{fig:global_homophilic}
{\bf Global homophilic percolation: staircase growth and fluctuations of the order parameter.}
10 realizations of the relative size of the largest cluster $S_1/N$ for the GHP process for $N=2^{30}$. The staircase-like growth clearly illustrates the strict size doubling mechanism and the occurrence of DSI.
Inset: The relative variance of the process for system sizes $N=2^{20}$, $N=2^{25}$ and $N=2^{30}$ for $R = 5000, 2500, 1500$ realizations, respectively. The critical point is marked by the largest peak with $p_c(N=2^{30}) \approx 0.629$. The localized peaks in the supercritical regime are the signature of the fixed positions of the doubling transitions and the DSI.
}
\end{figure}

Since only doubling transitions are possible the second largest cluster needs to grow before a merger with the largest cluster. This time leads to an intrinsic macroscopic delay between the growth steps of the largest cluster, leading to supercritical DSI. Typical realizations of the evolution of the relative size of the largest cluster are shown in Fig.~\ref{fig:global_homophilic} together with the relative variance of the order parameter.

Similar to models proposed in \cite{schrenk2012, nagler2012, nagler2013} this model features a continuous (first) percolation transition with an infinite amount of infinitely small discontinuous jumps. Remarkably, unlike the other models exhibiting this \enquote{staircase} behavior it is self-averaging. The positions of the peaks in the relative variance mark the individual steps after $p_c \approx 0.629$ (determined via $p_c \approx \mathrm{argmax}_p(R_v[S_1(p)])$, see Fig.~\ref{fig:global_homophilic}). These peaks localize for larger systems such that $R_v[S_1(p>p_c)] \rightarrow 0$ almost everywhere for $N \rightarrow \infty$.

This self-averaging (deterministic) staircase 
represents a new type of percolation phase transitions \cite{nagler2015} and 
is due to the fact that the discrete growth of clusters $S \rightarrow 2 S$ 
holds {\em always}, throughout the entire percolation process.
For the BFW-model on the lattice analyzed by Schrenk et al. \cite{schrenk2012} it remains to be explored if the observed staircase is a signature of non self-averaging such as in \cite{nagler2013}.\\

\begin{figure}[h]
\centering
\includegraphics[width = 0.45\textwidth]{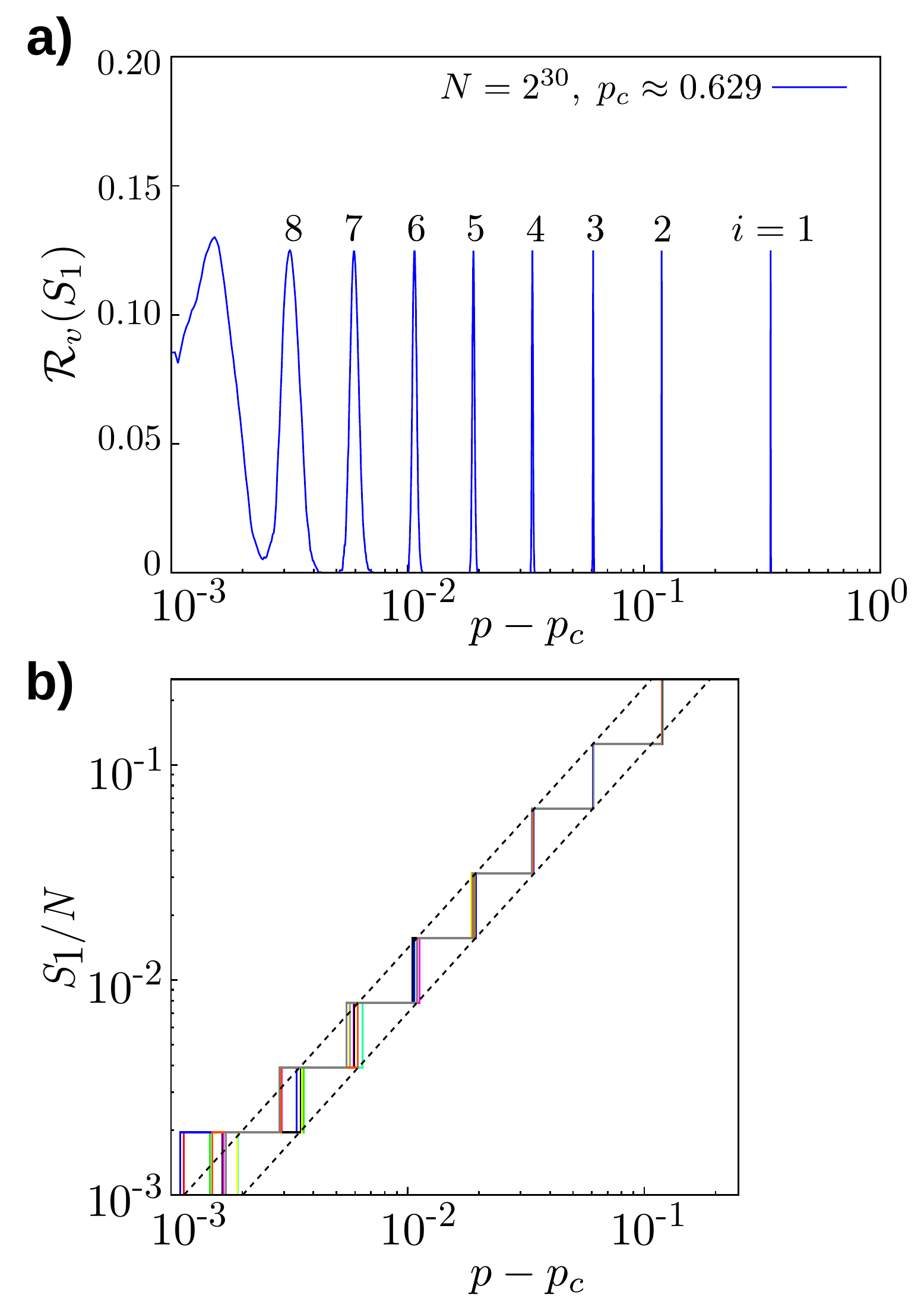}
\caption{
\label{fig:global_homophilic_DSI}
{\bf Global homophilic percolation: discrete scale invariance.}
\textbf{(a)} The relative variance for the GHP process with $N=2^{30}$ for $R=1500$ realizations in the supercritical regime, $p_c \approx 0.629$. In this presentation the log-periodicity, a hallmark of DSI, is clearly visible for intermediate values of $p-p_c$. \textbf{(b)} Sample realizations as in Fig.~\ref{fig:global_homophilic} together with the envelope (dashed lines) given by the estimated DSI scaling with exponent $\alpha \approx 1.19$ agreeing with the simulations. The intercept of the envelope was adjusted by hand.
}
\end{figure}

\textit{DSI in supercritical percolation}.---Fig.~\ref{fig:global_homophilic_DSI}(a) shows only the supercritical regime of the relative variance versus $p - p_c$. The delay between consecutive growth steps leads to positions of these steps that are log-periodic for intermediate values of $p - p_c$, signature of the DSI. Using Eq.~\ref{eq:dsi_notation} for pairs of consecutive steps we find an average $\lambda \approx 1.79$ using steps $3 \le i \le 6$. With the known factor $\mu = 2$ from the doubling transitions imposed by the percolation rule this leads to a scaling exponent of $\alpha \approx 1.19$. These results agree well with the actual evolution of the system as shown in Fig.~\ref{fig:global_homophilic_DSI}(b).

Due to the doubling of the size of the largest cluster with each step the number of observable jumps until the largest cluster reaches size $N$ scales as $\log(N)$ and thus the finite size simulations only show very few steps, even though in the thermodynamic limit the system exhibits infinitely many jumps close to $p_c$. In fact, it can be seen in Fig.~\ref{fig:global_homophilic_DSI}(a) that for very small values of $p - p_c$ the peaks are not log-periodic due to the finite size of the simulated systems.\\

\begin{figure}[h]
\centering
\includegraphics[width = 0.48\textwidth]{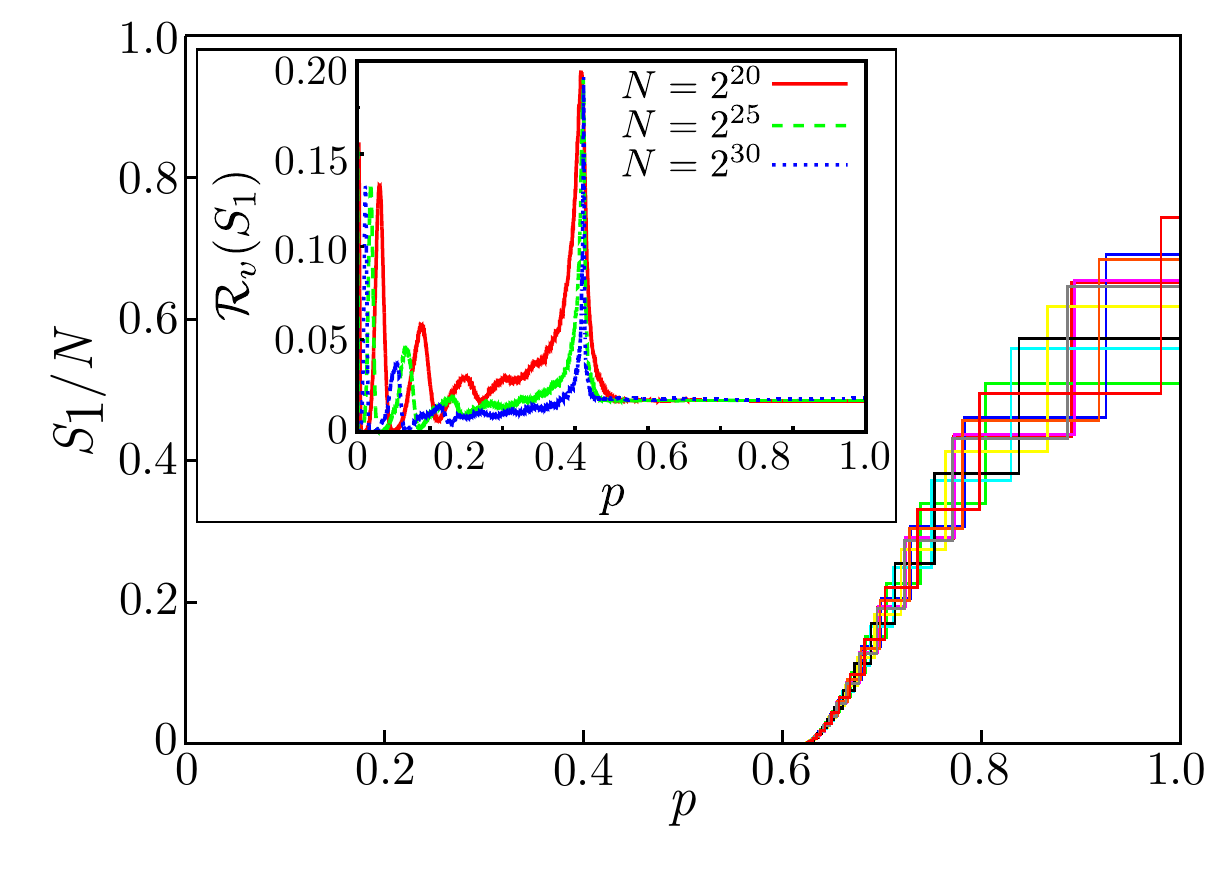}
\caption{
\label{fig:HP}
{\bf Homophilic percolation: no signature of DSI in the relative variance.}
10 realizations of the relative size of the largest cluster $S_1/N$ for the HP process for $N=2^{30}$. The process is non self-averaging: the positions of the discrete growth steps are randomly distributed and the realizations do not coincide. The inset shows the relative variance of the process for $N=2^{20}$, $N=2^{25}$ and $N=2^{30}$ for $R = 5000, 2500, 1500$ realizations respectively. The relative variance in the supercritical regime after $p_c \approx 0.6234$ (marked by the largest peak) does not decay to zero for larger systems indicating the non self-averaging. No evidence of log-periodic oscillation from DSI can be seen in this ensemble measurement.
}
\end{figure}

\textit{Homophilic percolation}.---A similar model favoring the merger of clusters of similar size was presented in \cite{nagler2012}.
In each step three nodes are chosen at random. Let the (ordered) sizes of the clusters they reside in be $S(1) \ge S(2) \ge S(3)$.
The two nodes that minimize $\delta_{i,j} = \left|S(i) - S(j)\right|$ are connected, possibly creating an intra-cluster link.
This homophilic percolation (HP) is the local competition variant of the global homophilic model discussed above. Example realizations of the process are shown in Fig.~\ref{fig:HP}.

Since the existence of a strict fractional growth mechanism $S_1\rightarrow \mu S_1$ is not obvious in this model we repeat the argument from \cite{nagler2012}:
Assume the largest cluster is the only macroscopic cluster, then it can never merge with another cluster.
Either two nodes in microscopic clusters are chosen and will be linked or two nodes from the largest cluster will be chosen and then linked.
Similarly, if two macroscopic clusters exist they can only merge if $S_2 \ge S_1 / 2$:
For such a merger the three nodes have to be drawn from the largest, second largest and a microscopic cluster, otherwise an intra-cluster link is added or two microscopic clusters merge.
Then the size difference $\delta_{2,3} = S(2) - S(3) = S_2 - o(N)$ between $S_2$ and a microscopic cluster of negligible size $o(N)$ must be larger than the difference $\delta_{1,2} = S(1) - S(2) = S_1-S_2$, requiring $S_2 \ge S_1 / 2$.
As soon as the cluster sizes allow such a merger it has a non-zero probability of occurring and will take place instantaneously (in $o(N)$ steps).
The HP-model therefore exhibits fractional growth with $\mu = 3/2$.\\

\textit{DSI in non self-averaging percolation}.---The relative variance in the HP-model (Fig.~\ref{fig:HP}) is non-zero after the percolation transition and does not disappear for larger systems indicating non self-averaging. This is also clearly visible by the fact that the different realizations do not coincide. Due to the non self-averaging no log-periodic oscillations and no signature of DSI is visible in the ensemble averaged properties such as the relative variance.

To clearly reveal the DSI even in the ensemble properties we propose the following rescaling procedure: In order to align the individual realizations we need to align both the relative size of the largest cluster as well as the link density. We do this by rescaling these quantities individually for each realization before calculating the ensemble properties, e.g., the relative variance. The rescaling is done in the following way:
\begin{eqnarray}
\frac{S_1(p)}{N} &\rightarrow & \frac{S_1^*(p)}{N} = k_S \cdot \frac{S_1(p)}{N}\,, \\
p - p_c &\rightarrow & \left(p-p_c\right)^* = k_p \cdot \left(p-p_c\right)\,,
\end{eqnarray}
where $k_S$ and $k_p$ are the rescaling factors depending on the realization. Specifically, we align the first step where $S_1(p_i)/N > 0.05$ such that $S_1^*(p_i)/N = 0.05$ and the following step such that it has a fixed length $(p_{i-1}-p_i)^* = 0.01$. Since the rescaling is linear it does not affect the scaling exponent close to $p_c$. The method is robust against changes in the exact values used for this rescaling procedure, however, the realizations should be aligned at a point where the DSI is expected to hold. Importantly, this rescaling does not affect the ensemble properties if the model is self-averaging as then the $k_S$ and $k_p$ are (almost) identical for all realizations.

\begin{figure}[h]
\centering
\includegraphics[width = 0.45\textwidth]{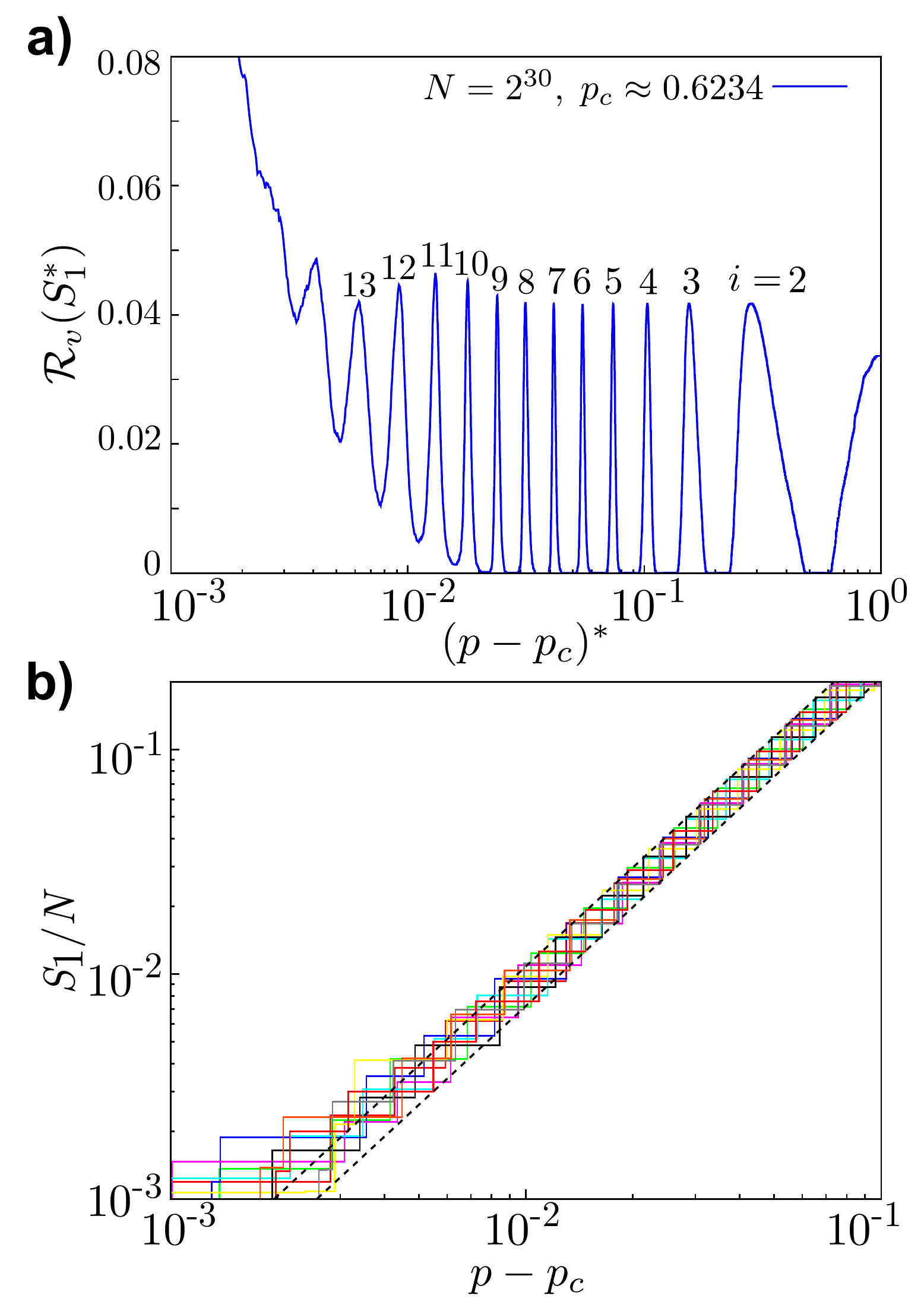}
\caption{
\label{fig:HP_DSI}
{\bf Homophilic percolation: revealing DSI in non self-averaging systems.}
\textbf{(a)} The relative variance for the HP-model with $N=2^{30}$ for $R=1500$ realizations in the supercritical regime, $p_c \approx 0.6234$, after applying the rescaling (see text). The rescaling aligns the individual realizations and leads to localized peaks in the relative variance marking the positions of the individual steps which show log-periodicity. Compared to Fig.~\ref{fig:HP} we can now estimate the DSI parameters $\lambda \approx 1.32$ from intermediate values of $p-p_c$.
\textbf{(b)} Sample realizations as in Fig.~\ref{fig:HP} together with the envelope (dashed lines) given by the estimated DSI exponent $\alpha \approx 1.46$ agreeing with the simulations. The intercept of the envelope was adjusted by hand.
}
\end{figure}

The effect of this rescaling on the relative variance is shown in Fig.~\ref{fig:HP_DSI}(a). The steps of the individual realizations are now localized at the same positions $\left(p-p_c\right)^*$ such that also the peaks in the relative variance become localized, similar to Fig.~\ref{fig:global_homophilic_DSI}.
This clearly reveals log-periodic oscillations and allows us to calculate the parameters of the DSI as above with $\lambda \approx 1.32$ (using $5 \le i \le 9$) and thus $\alpha \approx 1.46$ [see Fig.~\ref{fig:HP_DSI}(b)]. This shows that even though this model is non self-averaging and the positions of the growth steps are randomly distributed, the steps of a single realization are correlated via DSI. These correlations are such that once one observations of $S_1(p)/N$ is known the future (past) evolution can be perfectly predicted (reconstructed).\\

\textit{Modified ER-model}.---Another example for a model that shows supercritical DSI is the modified \ER{}-model (mER). This model was presented as one of the first examples of non self-averaging behavior in explosive percolation \cite{riordan2012} and is analytically treatable due to its similarities to standard ER-percolation \cite{ER}. It was pointed out that the non self-averaging in these models is caused by small fluctuations of $S_1$ close to the critical point that are then amplified by the fractional growth mechanism \cite{riordan2012}.
The mER-model restricts the largest (macroscopic) component to doubling transitions and we show here that it also features supercritical DSI.

Links are added as follows using the mER rule: 
Starting from an empty graph in each step three nodes 
are chosen at random.
If the two largest components in the entire network have the same size ($S_1 = S_2$), 
connect two of the chosen nodes at random. 
Otherwise ($S_1 > S_2$), 
 if at least two of the chosen nodes are inside the largest cluster those are connected.
Else two chosen nodes in the smaller clusters are connected. 
This model behaves exactly as the standard ER-model \cite{ER} for links that do not include the largest component.

\begin{figure}[h]
\centering
\includegraphics[width = 0.48\textwidth]{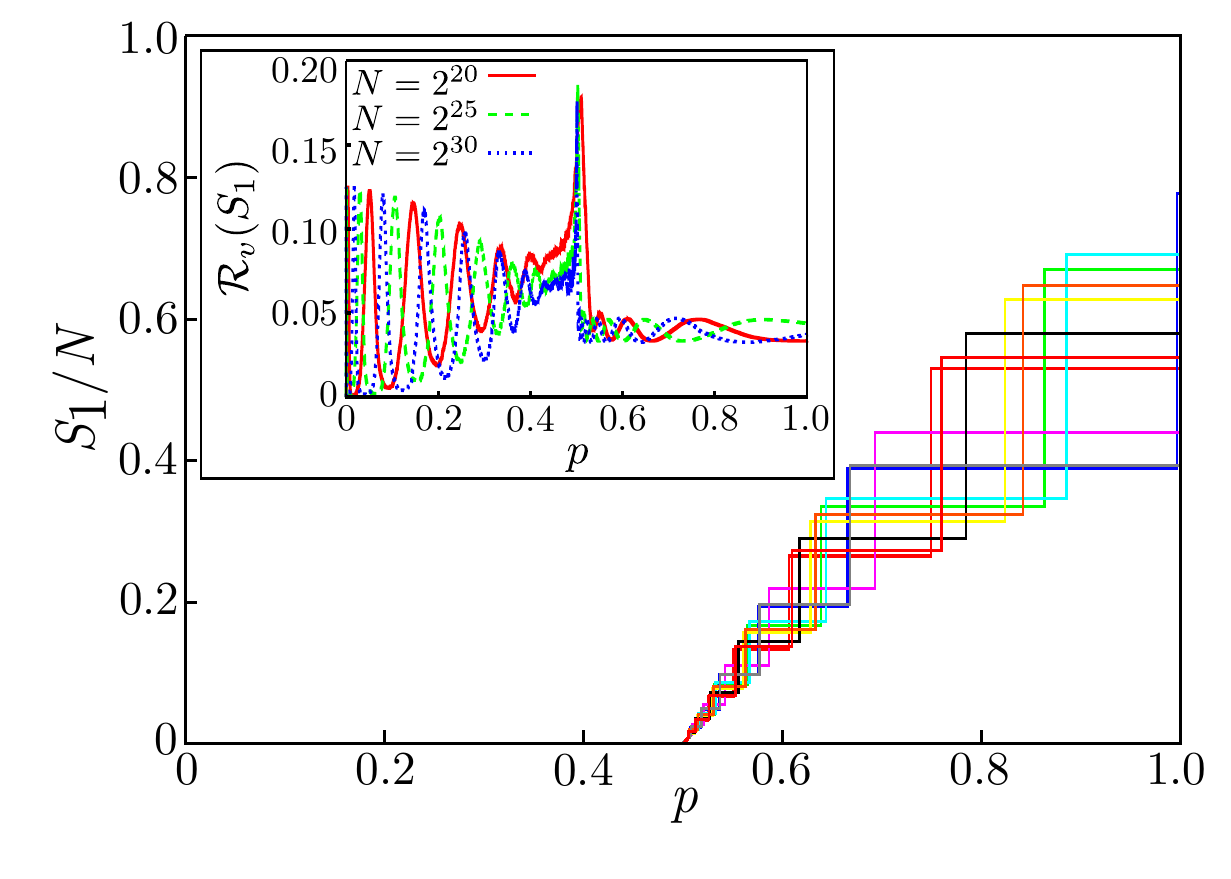}
\caption{
\label{fig:mER}
{\bf Modified ER-model: Non self-averaging blurs out log-periodic oscillations in the relative variance.}
10 realizations of the relative size of the largest cluster $S_1/N$ for the mER process for $N=2^{30}$. The process is non self-averaging: the positions of the discrete growth steps are randomly distributed. The inset shows the relative variance of the process for $N=2^{20}$, $N=2^{25}$ and $N=2^{30}$ for $R = 5000, 2500, 1500$ realizations, respectively. The relative variance in the supercritical regime after $p_c \approx 0.501$ (marked by the largest peak) does not decay to zero for larger systems indicating the non self-averaging. Due to the existence of preferred cluster sizes log-periodic oscillations are already visible as a signature of DSI but the peaks are blurred out due to non self-averaging of the model.
}
\end{figure}

Fig.~\ref{fig:mER} shows example realizations of the process and the relative variance of the largest cluster. The mER-model is non self-averaging evidenced by the non-zero relative variance in Fig~\ref{fig:mER}. However, there still exist preferred cluster sizes as can be seen by the small oscillations in $R_v[S_1(p)]$. Technically, even these peaks allow for calculating the DSI parameters $\mu$ and $\lambda$.
However, applying the rescaling as above reveals the DSI characteristics more clearly.

\begin{figure}[h]
\centering
\includegraphics[width = 0.45\textwidth]{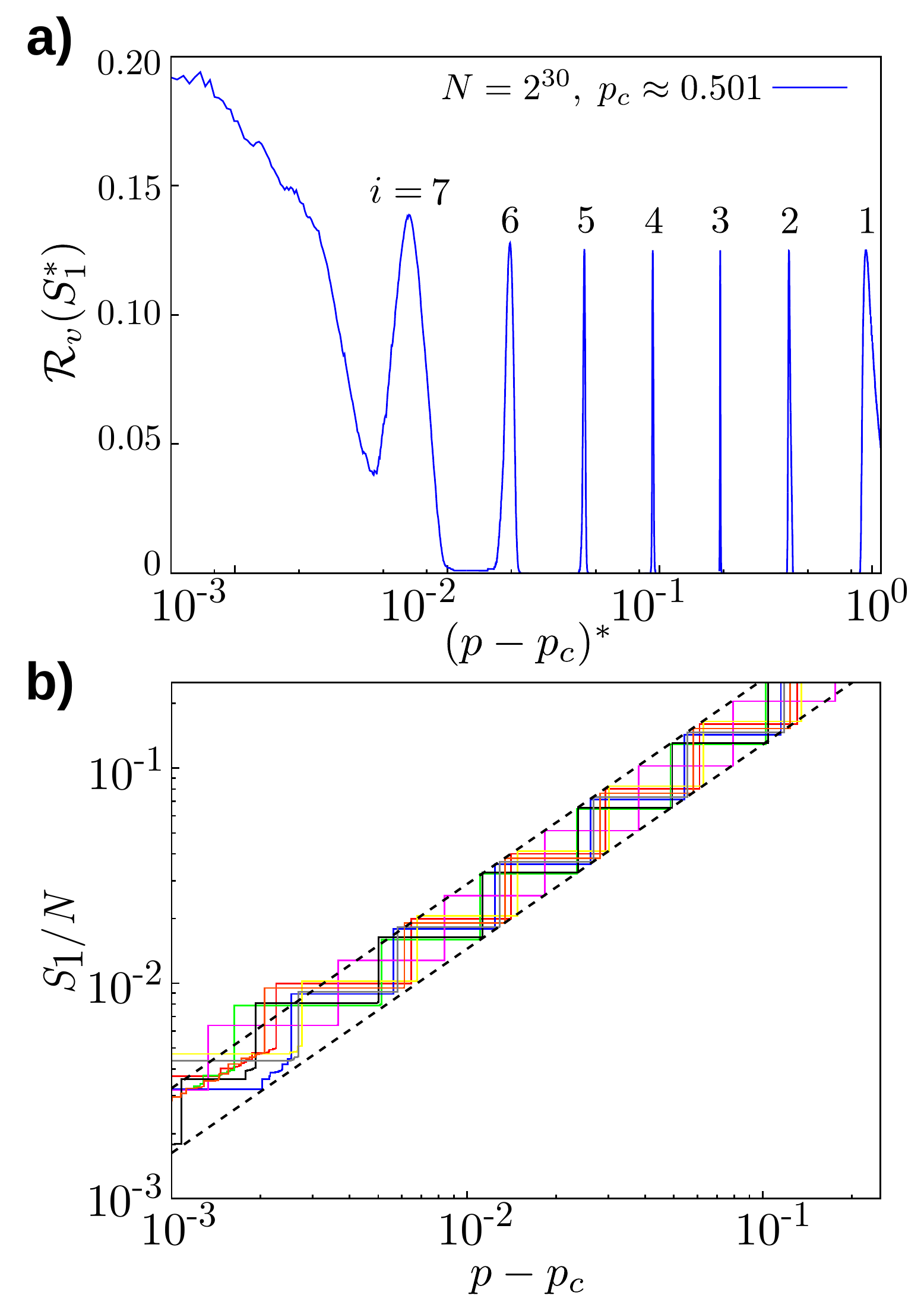}
\caption{
\label{fig:mER_DSI}
{\bf Modified ER-model: revealing DSI in non self-averaging systems.}
\textbf{(a)} The relative variance for the mER-model with $N=2^{30}$ for $R=1500$ realizations in the supercritical regime, $p_c \approx 0.501$, after applying the rescaling. The rescaling aligns the individual realizations and localizes the peaks in the relative variance marking the positions of the individual steps. Compared to Fig.~\ref{fig:mER} the log-periodicity is much clearer and we can estimate the DSI parameters $\lambda \approx 2.07$ from intermediate values of $p-p_c$.
\textbf{(b)} Sample realizations as in Fig.~\ref{fig:mER} together with the envelope (dashes lines) given by the estimated DSI exponent $\alpha \approx 0.95$ agreeing with the simulations. The intercept of the envelope was adjusted by hand.
}
\end{figure}

With the rescaling the steps of the individual realizations are now localized at the same positions such that the peaks in the relative variance become localized [Fig.~\ref{fig:mER_DSI}(a)] clearly revealing log-periodic oscillations and therefore the DSI. We again calculate the parameters of the DSI as above and find $\lambda \approx 2.07$ (using steps $2 \le i \le 5$) and thus $\alpha \approx 0.95$ [see Fig.~\ref{fig:mER_DSI}(b)]. Again, the small amount of observable jumps for finite system sizes decreases the possible accuracy of the measurement.\\

\textit{Analytic description of DSI}.---We can analytically explain the appearance of DSI in percolation models with a strict fractional growth mechanism. In particular, we can even calculate the scaling exponent in the thermodynamic limit for the mER-model.

In general we make two assumptions about the model: it exhibits strict fractional growth of the largest cluster $S_1(i) = \mu S_1(i+1)$ and power-law scaling (continuous scale invariance) close to the percolation transitions of the second largest cluster as is the case in most percolation models. Here, this means the delay between two consecutive jumps is given by $p_i - p_{i+1} = \Delta p_i \sim S_1(i+1) ^{1/\beta}$ while $S_2$ is growing to $S_2(p_i) = (\mu-1) S_1(i+1)$. We can then write $p_i = p_c + \sum_{j=i}^{\infty} \Delta p_j$ and with the two assumptions above we find
\begin{eqnarray}
p_i - p_c & \sim & \sum_{j=i}^{\infty} S_1(j+1) ^{1/\beta} = S_1(i+1) ^{1/\beta} \sum_{j=0}^{\infty} \frac{1}{\mu^{j/\beta}} \nonumber\\
	& = & \frac{\mu^{1/\beta}}{\mu^{1/\beta} - 1} S_1(i+1) ^{1/\beta} \sim S_1(i+1) ^{1/\beta} \,.
\end{eqnarray}
We therefore find as relation for the time between two consecutive steps
\begin{equation}
\lambda = \frac{p_i - p_c}{p_{i+1} - p_c} = \frac{S_1(i+1) ^{1/\beta}}{S_1(i+2) ^{1/\beta}} = \mu^{1/\beta} \,,
\end{equation}
giving the DSI exponent $\alpha = \beta$. This clearly shows the connection between DSI as an extension of the known scaling laws for $p> p_c$ and a strict fractional growth mechanism of the largest cluster.

For the mER-model these assumptions are fulfilled and we are in fact able to calculate the complete evolution of the largest cluster:
for all links not including the largest cluster the mER-model behaves exactly as a (link density rescaled) ER-model \cite{riordan2012}. Consider the time $\Delta p_i$ between two jumps of the largest cluster with size $S_1(i+1)$ as above.
$\Delta p_i$ is given by the time it takes for the second largest cluster to grow up to size $S_2(p_i) = S_1(i+1)$, comprised of growth in the subcritical and supercritical regime. The additional time needed due to the possibility of adding intra-cluster-links in the largest cluster (not affecting the growth of the second largest cluster) can be ignored for small $S_1/N$ close to $p_c$ as it only occurs with probability $\sim (S_1/N)^2$.
With $S_2(p)/N \approx 4 \cdot \left(p - p_{c,2}\right)$ for small $p-p_{c,2}$, known from the standard ER-model, it takes $\Delta p_i^\mathrm{super} \approx 1/4 \cdot S_1(i+1)/N$ time in the supercritical regime for $S_2$ to reach the size of $S_2(p_i) = S_1(i+1)$.
Due to the similarity of the sub- and supercritical regime in ER-percolation \cite{bollobas84} the time it takes for $S_2$ to reach the critical point from the last step is approximately the same as the time for the supercritical growth before the last step. Thus we have $\Delta p_i^\mathrm{sub} \approx \Delta p_{i+1}^\mathrm{super} \approx 1/4 \cdot S_1(i+2)/N = 1/8 \cdot S_1(i+1)/N$.

We then calculate the link density at the jump $i$ as above with $\mu = 2$ and $\beta = 1$
\begin{equation}
	p(i) - p_c = \sum_{j=0}^\infty\Delta p_j^\mathrm{sub}+\Delta p_j^\mathrm{super}  \approx \frac{3}{4} \frac{S_1(i+1)}{N}.
\end{equation}
We therefore find a DSI exponent $\alpha = 1$ ($\mu = 2$, $\lambda = 2$).
The growth of the largest cluster then follows a staircase with a lower envelope $\mathrm{min}(S_1(p)/N) \approx 4/3 \cdot (p-p_c)$ for $p$ close to $p_c$.
Both the exponent as well as the calculated prefactor agree well with the simulations considering the strong influence of finite size effects.\\

\textit{Beyond DSI in homophilic percolation}.--- Since we discussed two models of homophilic percolation as examples of (non) self-averaging percolation with supercritical DSI (see Fig.~\ref{fig:global_homophilic} and \ref{fig:HP}) one might expect similar results to hold for models in between the global and local homophilic percolation models. However, extending homophilic percolation to competition between more than $m=3$ nodes leads to much more complex behavior. In fact, both assumptions for the analytical derivation above are not given any more and the DSI is broken.

\begin{figure}[h]
\centering
\includegraphics[width = 0.48\textwidth]{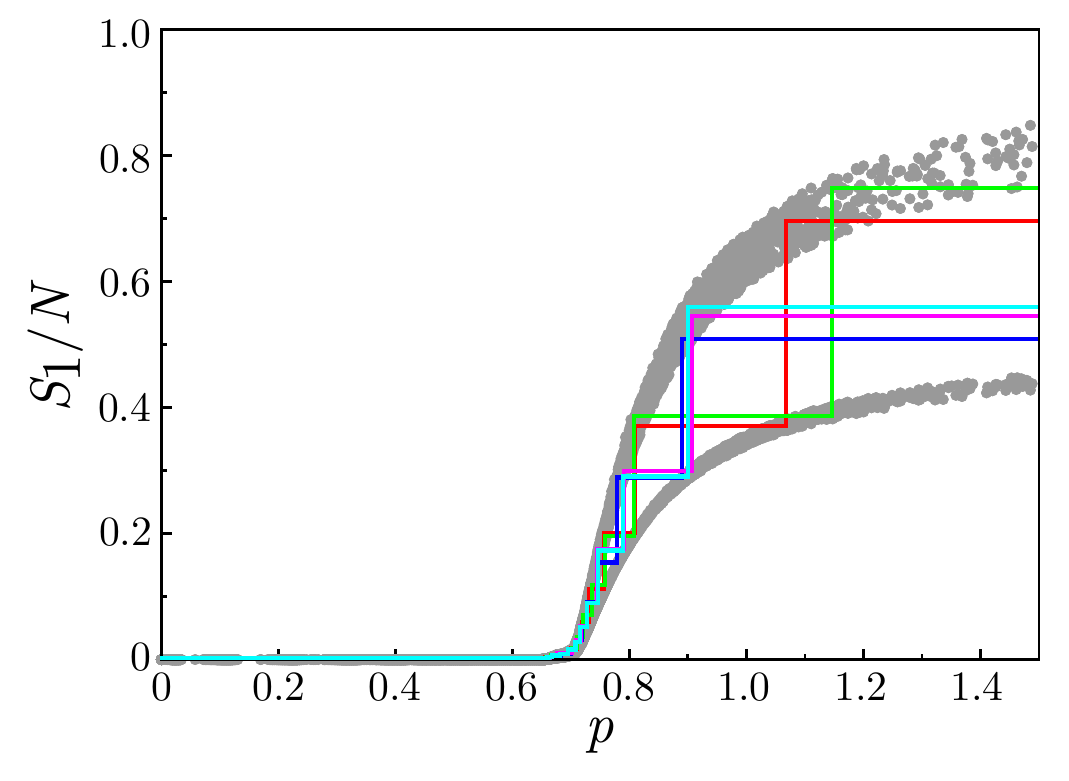}
\caption{
\label{fig:HP_m4}
{\bf Homophilic percolation: Extended competition breaks the discrete scale invariance.}
5 realizations of the relative size of the largest cluster $S_1/N$ for homophilic percolation with competition between $m=4$ nodes for $N=2^{30}$. The gray dots show the beginning and end points of steps from 1000 realizations showing a broad distribution. The process is non self-averaging but beginning and end points of the steps do not converge to well defined functions as in the case of $m=3$ homophilic percolation, the DSI is broken.
}
\end{figure}

For example, for $m=4$ homophilic percolation, i.e. choosing four nodes and connecting those that have the most similar cluster size, the \emph{second} largest cluster exhibits DSI after its percolation transitions. This leads to an additional source of randomness in the growth steps of the largest cluster. While this process is still non self-averaging, the randomness in the growth steps breaks the strict fractional growth of the largest cluster, breaking the discrete scale invariance (see Fig.~\ref{fig:HP_m4}). Competition between more nodes ($m>4$) leads to even more complex behavior as more sources of randomness are introduced. This shows that a \emph{strict} fractional growth mechanism is necessary for supercritical DSI in percolation.\\

\textit{Conclusion}.---We have studied percolation systems exhibiting a fractional growth mechanism, creating an intrinsic delay in the growth of the largest cluster, 
which leads 
to staircase-like growth and often non self-averaging. We demonstrated that this staircase growth extends the known power-law scaling in the supercritical regime to discrete scale invariance and analytically showed the connection between the scaling laws and the DSI parameters. For systems that are non self-averaging the signatures of DSI can be lost in ensemble-averaged properties. We proposed a rescaling method to align the individual realizations and expose the log-periodicity, the signature of DSI.

Notably, we introduced the
 global homophilic percolation model that exhibits a novel percolation phase transition type, characterized by a deterministic staircase \cite{nagler2015}.

In contrast to the previously discussed DSI in the subcritical regime \cite{nagler2014}, DSI in the supercritical regime of percolation cannot stem from some underlying discreteness of the network or lattice. We have demonstrated the existence of a strict fractional growth mechanism of the largest cluster as a condition for supercritical DSI. This fractional growth leads to an implicit delay in the growth of the largest cluster leading to DSI.
Conversely, it would be interesting to see whether DSI can also be induced by an explicit delay in percolation models, especially since link addition or the merger of clusters in real networks is often not instantaneous.

\begin{acknowledgments}
Malte Schr\"oder acknowledges support from the Göttingen Graduate School for Neurosciences and Molecular Biosciences (DFG Grant GSC 226/2).
Wei Chen would like to acknowledge support from the National Natural Science Foundation of China under Grant No. 11305219.
Jan Nagler acknowledges the ETH Risk Center for support (Grant No. SP RC 08-15).
\end{acknowledgments}

\bibliography{manuscript_supercriticalDSI}

\begin{thebibliography}{31}%
\makeatletter
\providecommand \@ifxundefined [1]{%
 \@ifx{#1\undefined}
}%
\providecommand \@ifnum [1]{%
 \ifnum #1\expandafter \@firstoftwo
 \else \expandafter \@secondoftwo
 \fi
}%
\providecommand \@ifx [1]{%
 \ifx #1\expandafter \@firstoftwo
 \else \expandafter \@secondoftwo
 \fi
}%
\providecommand \natexlab [1]{#1}%
\providecommand \enquote  [1]{``#1''}%
\providecommand \bibnamefont  [1]{#1}%
\providecommand \bibfnamefont [1]{#1}%
\providecommand \citenamefont [1]{#1}%
\providecommand \href@noop [0]{\@secondoftwo}%
\providecommand \href [0]{\begingroup \@sanitize@url \@href}%
\providecommand \@href[1]{\@@startlink{#1}\@@href}%
\providecommand \@@href[1]{\endgroup#1\@@endlink}%
\providecommand \@sanitize@url [0]{\catcode `\\12\catcode `\$12\catcode
  `\&12\catcode `\#12\catcode `\^12\catcode `\_12\catcode `\%12\relax}%
\providecommand \@@startlink[1]{}%
\providecommand \@@endlink[0]{}%
\providecommand \url  [0]{\begingroup\@sanitize@url \@url }%
\providecommand \@url [1]{\endgroup\@href {#1}{\urlprefix }}%
\providecommand \urlprefix  [0]{URL }%
\providecommand \Eprint [0]{\href }%
\@ifxundefined \urlstyle {%
  \providecommand \doi  [0]{\begingroup \@sanitize@url \@doi}%
  \providecommand \@doi [1]{\endgroup \@@startlink {\doibase
  #1}doi:\discretionary {}{}{}#1\@@endlink }%
}{%
  \providecommand \doi  [0]{doi:\discretionary{}{}{}\begingroup
  \urlstyle{rm}\Url }%
}%
\providecommand \doibase [0]{http://dx.doi.org/}%
\providecommand \Doi [0]{\begingroup \@sanitize@url \@Doi }%
\providecommand \@Doi  [1]{\endgroup\@@startlink{\doibase#1}\@@Doi}%
\providecommand \@@Doi [1]{#1\@@endlink}%
\providecommand \selectlanguage [0]{\@gobble}%
\providecommand \bibinfo  [0]{\@secondoftwo}%
\providecommand \bibfield  [0]{\@secondoftwo}%
\providecommand \translation [1]{[#1]}%
\providecommand \BibitemOpen [0]{}%
\providecommand \bibitemStop [0]{}%
\providecommand \bibitemNoStop [0]{.\EOS\space}%
\providecommand \EOS [0]{\spacefactor3000\relax}%
\providecommand \BibitemShut  [1]{\csname bibitem#1\endcsname}%
\bibitem [{\citenamefont {Stauffer}\ and\ \citenamefont
  {Aharony}(1994)}]{StaufferPercBook}%
  \BibitemOpen
  \bibfield  {author} {\bibinfo {author} {\bibfnamefont {D.}~\bibnamefont
  {Stauffer}}\ and\ \bibinfo {author} {\bibfnamefont {A.}~\bibnamefont
  {Aharony}},\ }\href@noop {} {\emph {\bibinfo {title} {Introduction To
  Percolation Theory: Revised Second Edition}}}\ (\bibinfo  {publisher} {Taylor
  \& Francis},\ \bibinfo {year} {1994})\BibitemShut {NoStop}%
\bibitem [{\citenamefont {Gaillard-Groleas}\ \emph {et~al.}(1990)\citenamefont
  {Gaillard-Groleas}, \citenamefont {Lagier},\ and\ \citenamefont
  {Sornette}}]{Sornette1990}%
  \BibitemOpen
  \bibfield  {author} {\bibinfo {author} {\bibfnamefont {G.}~\bibnamefont
  {Gaillard-Groleas}}, \bibinfo {author} {\bibfnamefont {M.}~\bibnamefont
  {Lagier}}, \ and\ \bibinfo {author} {\bibfnamefont {D.}~\bibnamefont
  {Sornette}},\ }\href@noop {} {\bibfield  {journal} {\bibinfo  {journal}
  {Phys. Rev. Lett.},\ }\textbf {\bibinfo {volume} {64}},\ \bibinfo {pages}
  {1577} (\bibinfo {year} {1990})}\BibitemShut {NoStop}%
\bibitem [{\citenamefont {Drossel}\ and\ \citenamefont
  {Schwabl}(1992)}]{DrosselPRL1992}%
  \BibitemOpen
  \bibfield  {author} {\bibinfo {author} {\bibfnamefont {B.}~\bibnamefont
  {Drossel}}\ and\ \bibinfo {author} {\bibfnamefont {F.}~\bibnamefont
  {Schwabl}},\ }\href@noop {} {\bibfield  {journal} {\bibinfo  {journal} {Phys.
  Rev. Lett.},\ }\textbf {\bibinfo {volume} {69}},\ \bibinfo {pages} {1629}
  (\bibinfo {year} {1992})}\BibitemShut {NoStop}%
\bibitem [{\citenamefont {Buldyrev}\ \emph {et~al.}(2010)\citenamefont
  {Buldyrev}, \citenamefont {Parshani}, \citenamefont {Paul}, \citenamefont
  {Stanley},\ and\ \citenamefont {Havlin}}]{ParshaniandBulyrevandStanley2010}%
  \BibitemOpen
  \bibfield  {author} {\bibinfo {author} {\bibfnamefont {S.~V.}\ \bibnamefont
  {Buldyrev}}, \bibinfo {author} {\bibfnamefont {R.}~\bibnamefont {Parshani}},
  \bibinfo {author} {\bibfnamefont {G.}~\bibnamefont {Paul}}, \bibinfo {author}
  {\bibfnamefont {H.~E.}\ \bibnamefont {Stanley}}, \ and\ \bibinfo {author}
  {\bibfnamefont {S.}~\bibnamefont {Havlin}},\ }\href@noop {} {\bibfield
  {journal} {\bibinfo  {journal} {Nature},\ }\textbf {\bibinfo {volume}
  {464}},\ \bibinfo {pages} {1025} (\bibinfo {year} {2010})}\BibitemShut
  {NoStop}%
\bibitem [{\citenamefont {Callaway}\ \emph {et~al.}(2000)\citenamefont
  {Callaway}, \citenamefont {Newman}, \citenamefont {Strogatz},\ and\
  \citenamefont {Watts}}]{CallawayPRL2000}%
  \BibitemOpen
  \bibfield  {author} {\bibinfo {author} {\bibfnamefont {D.~S.}\ \bibnamefont
  {Callaway}}, \bibinfo {author} {\bibfnamefont {M.~E.~J.}\ \bibnamefont
  {Newman}}, \bibinfo {author} {\bibfnamefont {S.~H.}\ \bibnamefont
  {Strogatz}}, \ and\ \bibinfo {author} {\bibfnamefont {D.~J.}\ \bibnamefont
  {Watts}},\ }\href@noop {} {\bibfield  {journal} {\bibinfo  {journal} {Phys.
  Rev. Lett.},\ }\textbf {\bibinfo {volume} {85}},\ \bibinfo {pages} {5468}
  (\bibinfo {year} {2000})}\BibitemShut {NoStop}%
\bibitem [{\citenamefont {Newman}\ \emph {et~al.}(2002)\citenamefont {Newman},
  \citenamefont {Watts},\ and\ \citenamefont {Strogatz}}]{NewmanandWatts2002}%
  \BibitemOpen
  \bibfield  {author} {\bibinfo {author} {\bibfnamefont {M.~E.~J.}\
  \bibnamefont {Newman}}, \bibinfo {author} {\bibfnamefont {D.~J.}\
  \bibnamefont {Watts}}, \ and\ \bibinfo {author} {\bibfnamefont {S.~H.}\
  \bibnamefont {Strogatz}},\ }\href@noop {} {\bibfield  {journal} {\bibinfo
  {journal} {Proc. Natl. Acad. Sci.},\ }\textbf {\bibinfo {volume} {99}},\
  \bibinfo {pages} {2566} (\bibinfo {year} {2002})}\BibitemShut {NoStop}%
\bibitem [{\citenamefont {Ali~Saberi}(2013)}]{saberi13}%
  \BibitemOpen
  \bibfield  {author} {\bibinfo {author} {\bibfnamefont {A.}~\bibnamefont
  {Ali~Saberi}},\ }\href@noop {} {\bibfield  {journal} {\bibinfo  {journal}
  {Phys. Rev. Lett.},\ }\textbf {\bibinfo {volume} {110}},\ \bibinfo {pages}
  {178501} (\bibinfo {year} {2013})}\BibitemShut {NoStop}%
\bibitem [{\citenamefont {D'Souza}\ and\ \citenamefont
  {Nagler}(2015)}]{nagler2015}%
  \BibitemOpen
  \bibfield  {author} {\bibinfo {author} {\bibfnamefont {R.~M.}\ \bibnamefont
  {D'Souza}}\ and\ \bibinfo {author} {\bibfnamefont {J.}~\bibnamefont
  {Nagler}},\ }\href@noop {} {\bibfield  {journal} {\bibinfo  {journal} {Nat.
  Phys.},\ }\textbf {\bibinfo {volume} {11}},\ \bibinfo {pages} {531} (\bibinfo
  {year} {2015})}\BibitemShut {NoStop}%
\bibitem [{\citenamefont {Sahini}\ and\ \citenamefont {Sahimi}(1994)}]{Sahimi}%
  \BibitemOpen
  \bibfield  {author} {\bibinfo {author} {\bibfnamefont {M.}~\bibnamefont
  {Sahini}}\ and\ \bibinfo {author} {\bibfnamefont {M.}~\bibnamefont
  {Sahimi}},\ }\href@noop {} {\emph {\bibinfo {title} {Applications Of
  Percolation Theory}}}\ (\bibinfo  {publisher} {CRC Press},\ \bibinfo {year}
  {1994})\BibitemShut {NoStop}%
\bibitem [{\citenamefont {Andrade~Jr}\ \emph {et~al.}(2000)\citenamefont
  {Andrade~Jr}, \citenamefont {Buldyrev}, \citenamefont {Dokholyan},
  \citenamefont {Havlin}, \citenamefont {King}, \citenamefont {Lee},
  \citenamefont {Paul},\ and\ \citenamefont {Stanley}}]{AndradePRE2000}%
  \BibitemOpen
  \bibfield  {author} {\bibinfo {author} {\bibfnamefont {J.~S.}\ \bibnamefont
  {Andrade~Jr}}, \bibinfo {author} {\bibfnamefont {S.~V.}\ \bibnamefont
  {Buldyrev}}, \bibinfo {author} {\bibfnamefont {N.~V.}\ \bibnamefont
  {Dokholyan}}, \bibinfo {author} {\bibfnamefont {S.}~\bibnamefont {Havlin}},
  \bibinfo {author} {\bibfnamefont {P.~R.}\ \bibnamefont {King}}, \bibinfo
  {author} {\bibfnamefont {Y.}~\bibnamefont {Lee}}, \bibinfo {author}
  {\bibfnamefont {G.}~\bibnamefont {Paul}}, \ and\ \bibinfo {author}
  {\bibfnamefont {H.~E.}\ \bibnamefont {Stanley}},\ }\href@noop {} {\bibfield
  {journal} {\bibinfo  {journal} {Phys. Rev. E},\ }\textbf {\bibinfo {volume}
  {62}},\ \bibinfo {pages} {8270} (\bibinfo {year} {2000})}\BibitemShut
  {NoStop}%
\bibitem [{\citenamefont {Ziff}\ \emph {et~al.}(1982)\citenamefont {Ziff},
  \citenamefont {Hendriks},\ and\ \citenamefont {Ernst}}]{ZiffPRL1982}%
  \BibitemOpen
  \bibfield  {author} {\bibinfo {author} {\bibfnamefont {R.~M.}\ \bibnamefont
  {Ziff}}, \bibinfo {author} {\bibfnamefont {E.~M.}\ \bibnamefont {Hendriks}},
  \ and\ \bibinfo {author} {\bibfnamefont {M.~H.}\ \bibnamefont {Ernst}},\
  }\href@noop {} {\bibfield  {journal} {\bibinfo  {journal} {Phys. Rev.
  Lett.},\ }\textbf {\bibinfo {volume} {49}},\ \bibinfo {pages} {593} (\bibinfo
  {year} {1982})}\BibitemShut {NoStop}%
\bibitem [{\citenamefont {Anderson}\ \emph {et~al.}(1992)\citenamefont
  {Anderson}, \citenamefont {May},\ and\ \citenamefont
  {Anderson}}]{Anderson1991}%
  \BibitemOpen
  \bibfield  {author} {\bibinfo {author} {\bibfnamefont {R.~M.}\ \bibnamefont
  {Anderson}}, \bibinfo {author} {\bibfnamefont {R.~M.}\ \bibnamefont {May}}, \
  and\ \bibinfo {author} {\bibfnamefont {B.}~\bibnamefont {Anderson}},\
  }\href@noop {} {\emph {\bibinfo {title} {Infectious diseases of humans:
  dynamics and control}}},\ Vol.~\bibinfo {volume} {28}\ (\bibinfo  {publisher}
  {Oxford University Press},\ \bibinfo {year} {1992})\BibitemShut {NoStop}%
\bibitem [{\citenamefont {Moore}\ and\ \citenamefont
  {Newman}(2000)}]{CmoorePRE2000}%
  \BibitemOpen
  \bibfield  {author} {\bibinfo {author} {\bibfnamefont {C.}~\bibnamefont
  {Moore}}\ and\ \bibinfo {author} {\bibfnamefont {M.~E.~J.}\ \bibnamefont
  {Newman}},\ }\href@noop {} {\bibfield  {journal} {\bibinfo  {journal} {Phys.
  Rev. E},\ }\textbf {\bibinfo {volume} {61}},\ \bibinfo {pages} {5678}
  (\bibinfo {year} {2000})}\BibitemShut {NoStop}%
\bibitem [{\citenamefont {Pastor-Satorras}\ and\ \citenamefont
  {Vespignani}(2001)}]{PastorPRL2001}%
  \BibitemOpen
  \bibfield  {author} {\bibinfo {author} {\bibfnamefont {R.}~\bibnamefont
  {Pastor-Satorras}}\ and\ \bibinfo {author} {\bibfnamefont {A.}~\bibnamefont
  {Vespignani}},\ }\href@noop {} {\bibfield  {journal} {\bibinfo  {journal}
  {Phys. Rev. Lett.},\ }\textbf {\bibinfo {volume} {86}},\ \bibinfo {pages}
  {3200} (\bibinfo {year} {2001})}\BibitemShut {NoStop}%
\bibitem [{\citenamefont {Strang}\ and\ \citenamefont
  {Soule}(1998)}]{StrangARS1998}%
  \BibitemOpen
  \bibfield  {author} {\bibinfo {author} {\bibfnamefont {D.}~\bibnamefont
  {Strang}}\ and\ \bibinfo {author} {\bibfnamefont {S.~A.}\ \bibnamefont
  {Soule}},\ }\href@noop {} {\bibfield  {journal} {\bibinfo  {journal} {Annu.
  Rev. Sociol.},\ }\textbf {\bibinfo {volume} {24}},\ \bibinfo {pages} {265}
  (\bibinfo {year} {1998})}\BibitemShut {NoStop}%
\bibitem [{\citenamefont {Lazarsfeld}\ \emph {et~al.}(1944)\citenamefont
  {Lazarsfeld}, \citenamefont {Berelson},\ and\ \citenamefont
  {Gaudet}}]{Lazarsfeld1944}%
  \BibitemOpen
  \bibfield  {author} {\bibinfo {author} {\bibfnamefont {P.~F.}\ \bibnamefont
  {Lazarsfeld}}, \bibinfo {author} {\bibfnamefont {B.}~\bibnamefont
  {Berelson}}, \ and\ \bibinfo {author} {\bibfnamefont {H.}~\bibnamefont
  {Gaudet}},\ }\href@noop {} {\emph {\bibinfo {title} {The People's Choice}}}\
  (\bibinfo  {publisher} {Columbia University Press},\ \bibinfo {year}
  {1944})\BibitemShut {NoStop}%
\bibitem [{\citenamefont {Stanley}(1971)}]{stanley71}%
  \BibitemOpen
  \bibfield  {author} {\bibinfo {author} {\bibfnamefont {H.~E.}\ \bibnamefont
  {Stanley}},\ }\href@noop {} {\emph {\bibinfo {title} {Introduction to Phase
  Transitions and Critical Phenomena}}},\ The International Series of
  Monographs on Physics\ (\bibinfo  {publisher} {Oxford University Press},\
  \bibinfo {year} {1971})\BibitemShut {NoStop}%
\bibitem [{\citenamefont {Sornette}(2004)}]{sornettebook}%
  \BibitemOpen
  \bibfield  {author} {\bibinfo {author} {\bibfnamefont {D.}~\bibnamefont
  {Sornette}},\ }\href@noop {} {\emph {\bibinfo {title} {Critical Phenomena in
  Natural Sciences, Chaos, Fractals, Self-organization and Disorder: Concepts
  and Tools}}},\ Springer Series in Synergetics\ (\bibinfo  {publisher}
  {Springer-Verlag Berlin Heidelberg},\ \bibinfo {year} {2004})\BibitemShut
  {NoStop}%
\bibitem [{\citenamefont {Sornette}(1998)}]{sornetteDSI}%
  \BibitemOpen
  \bibfield  {author} {\bibinfo {author} {\bibfnamefont {D.}~\bibnamefont
  {Sornette}},\ }\href@noop {} {\bibfield  {journal} {\bibinfo  {journal}
  {Phys. Rep.},\ }\textbf {\bibinfo {volume} {297}},\ \bibinfo {pages} {239}
  (\bibinfo {year} {1998})}\BibitemShut {NoStop}%
\bibitem [{\citenamefont {Chen}\ \emph {et~al.}(2014)\citenamefont {Chen},
  \citenamefont {Schr{\"o}der}, \citenamefont {D'Souza}, \citenamefont
  {Sornette},\ and\ \citenamefont {Nagler}}]{nagler2014}%
  \BibitemOpen
  \bibfield  {author} {\bibinfo {author} {\bibfnamefont {W.}~\bibnamefont
  {Chen}}, \bibinfo {author} {\bibfnamefont {M.}~\bibnamefont {Schr{\"o}der}},
  \bibinfo {author} {\bibfnamefont {R.~M.}\ \bibnamefont {D'Souza}}, \bibinfo
  {author} {\bibfnamefont {D.}~\bibnamefont {Sornette}}, \ and\ \bibinfo
  {author} {\bibfnamefont {J.}~\bibnamefont {Nagler}},\ }\href@noop {}
  {\bibfield  {journal} {\bibinfo  {journal} {Phys. Rev. Lett.},\ }\textbf
  {\bibinfo {volume} {112}},\ \bibinfo {pages} {155701} (\bibinfo {year}
  {2014})}\BibitemShut {NoStop}%
\bibitem [{\citenamefont {Chen}\ \emph
  {et~al.}(2013){\natexlab{a}}\citenamefont {Chen}, \citenamefont {Cheng},
  \citenamefont {Zheng}, \citenamefont {Chung}, \citenamefont {D'Souza},\ and\
  \citenamefont {Nagler}}]{chen13unstable}%
  \BibitemOpen
  \bibfield  {author} {\bibinfo {author} {\bibfnamefont {W.}~\bibnamefont
  {Chen}}, \bibinfo {author} {\bibfnamefont {X.}~\bibnamefont {Cheng}},
  \bibinfo {author} {\bibfnamefont {Z.}~\bibnamefont {Zheng}}, \bibinfo
  {author} {\bibfnamefont {N.~N.}\ \bibnamefont {Chung}}, \bibinfo {author}
  {\bibfnamefont {R.~M.}\ \bibnamefont {D'Souza}}, \ and\ \bibinfo {author}
  {\bibfnamefont {J.}~\bibnamefont {Nagler}},\ }\href@noop {} {\bibfield
  {journal} {\bibinfo  {journal} {Phys. Rev. E},\ }\textbf {\bibinfo {volume}
  {88}},\ \bibinfo {pages} {042152} (\bibinfo {year}
  {2013}{\natexlab{a}})}\BibitemShut {NoStop}%
\bibitem [{\citenamefont {Chen}\ \emph
  {et~al.}(2013){\natexlab{b}}\citenamefont {Chen}, \citenamefont {Nagler},
  \citenamefont {Cheng}, \citenamefont {Jin}, \citenamefont {Shen},
  \citenamefont {Zheng},\ and\ \citenamefont {D'Souza}}]{chen2013phase}%
  \BibitemOpen
  \bibfield  {author} {\bibinfo {author} {\bibfnamefont {W.}~\bibnamefont
  {Chen}}, \bibinfo {author} {\bibfnamefont {J.}~\bibnamefont {Nagler}},
  \bibinfo {author} {\bibfnamefont {X.}~\bibnamefont {Cheng}}, \bibinfo
  {author} {\bibfnamefont {X.}~\bibnamefont {Jin}}, \bibinfo {author}
  {\bibfnamefont {H.}~\bibnamefont {Shen}}, \bibinfo {author} {\bibfnamefont
  {Z.}~\bibnamefont {Zheng}}, \ and\ \bibinfo {author} {\bibfnamefont {R.~M.}\
  \bibnamefont {D'Souza}},\ }\href@noop {} {\bibfield  {journal} {\bibinfo
  {journal} {Phys. Rev. E},\ }\textbf {\bibinfo {volume} {87}},\ \bibinfo
  {pages} {052130} (\bibinfo {year} {2013}{\natexlab{b}})}\BibitemShut
  {NoStop}%
\bibitem [{\citenamefont {Nagler}\ \emph {et~al.}(2012)\citenamefont {Nagler},
  \citenamefont {Tiessen},\ and\ \citenamefont {Gutch}}]{nagler2012}%
  \BibitemOpen
  \bibfield  {author} {\bibinfo {author} {\bibfnamefont {J.}~\bibnamefont
  {Nagler}}, \bibinfo {author} {\bibfnamefont {T.}~\bibnamefont {Tiessen}}, \
  and\ \bibinfo {author} {\bibfnamefont {H.~W.}\ \bibnamefont {Gutch}},\
  }\href@noop {} {\bibfield  {journal} {\bibinfo  {journal} {Phys. Rev. X},\
  }\textbf {\bibinfo {volume} {2}},\ \bibinfo {pages} {031009} (\bibinfo {year}
  {2012})}\BibitemShut {NoStop}%
\bibitem [{\citenamefont {Schr{\"o}der}\ \emph {et~al.}(2013)\citenamefont
  {Schr{\"o}der}, \citenamefont {Rahbari},\ and\ \citenamefont
  {Nagler}}]{nagler2013}%
  \BibitemOpen
  \bibfield  {author} {\bibinfo {author} {\bibfnamefont {M.}~\bibnamefont
  {Schr{\"o}der}}, \bibinfo {author} {\bibfnamefont {S.~H.~E.}\ \bibnamefont
  {Rahbari}}, \ and\ \bibinfo {author} {\bibfnamefont {J.}~\bibnamefont
  {Nagler}},\ }\href@noop {} {\bibfield  {journal} {\bibinfo  {journal} {Nat.
  Commun.},\ }\textbf {\bibinfo {volume} {4}} (\bibinfo {year}
  {2013})}\BibitemShut {NoStop}%
\bibitem [{\citenamefont {Riordan}\ and\ \citenamefont
  {Warnke}(2012)}]{riordan2012}%
  \BibitemOpen
  \bibfield  {author} {\bibinfo {author} {\bibfnamefont {O.}~\bibnamefont
  {Riordan}}\ and\ \bibinfo {author} {\bibfnamefont {L.}~\bibnamefont
  {Warnke}},\ }\href@noop {} {\bibfield  {journal} {\bibinfo  {journal} {Phys.
  Rev. E},\ }\textbf {\bibinfo {volume} {86}},\ \bibinfo {pages} {011129}
  (\bibinfo {year} {2012})}\BibitemShut {NoStop}%
\bibitem [{\citenamefont {Achlioptas}\ \emph {et~al.}(2009)\citenamefont
  {Achlioptas}, \citenamefont {D'Souza},\ and\ \citenamefont
  {Spencer}}]{achlioptas}%
  \BibitemOpen
  \bibfield  {author} {\bibinfo {author} {\bibfnamefont {D.}~\bibnamefont
  {Achlioptas}}, \bibinfo {author} {\bibfnamefont {R.}~\bibnamefont {D'Souza}},
  \ and\ \bibinfo {author} {\bibfnamefont {J.}~\bibnamefont {Spencer}},\
  }\href@noop {} {\bibfield  {journal} {\bibinfo  {journal} {Science},\
  }\textbf {\bibinfo {volume} {323}},\ \bibinfo {pages} {1453} (\bibinfo {year}
  {2009})}\BibitemShut {NoStop}%
\bibitem [{\citenamefont {Friedman}\ and\ \citenamefont
  {Landsberg}(2009)}]{friedman2009}%
  \BibitemOpen
  \bibfield  {author} {\bibinfo {author} {\bibfnamefont {E.~J.}\ \bibnamefont
  {Friedman}}\ and\ \bibinfo {author} {\bibfnamefont {A.~S.}\ \bibnamefont
  {Landsberg}},\ }\href@noop {} {\bibfield  {journal} {\bibinfo  {journal}
  {Phys. Rev. Lett.},\ }\textbf {\bibinfo {volume} {103}},\ \bibinfo {pages}
  {255701} (\bibinfo {year} {2009})}\BibitemShut {NoStop}%
\bibitem [{\citenamefont {Nagler}\ \emph {et~al.}(2011)\citenamefont {Nagler},
  \citenamefont {Levina},\ and\ \citenamefont {Timme}}]{nagler2011}%
  \BibitemOpen
  \bibfield  {author} {\bibinfo {author} {\bibfnamefont {J.}~\bibnamefont
  {Nagler}}, \bibinfo {author} {\bibfnamefont {A.}~\bibnamefont {Levina}}, \
  and\ \bibinfo {author} {\bibfnamefont {M.}~\bibnamefont {Timme}},\
  }\href@noop {} {\bibfield  {journal} {\bibinfo  {journal} {Nat. Phys.},\
  }\textbf {\bibinfo {volume} {7}},\ \bibinfo {pages} {265} (\bibinfo {year}
  {2011})}\BibitemShut {NoStop}%
\bibitem [{\citenamefont {Schrenk}\ \emph {et~al.}(2012)\citenamefont
  {Schrenk}, \citenamefont {Felder}, \citenamefont {Deflorin}, \citenamefont
  {Ara\'ujo}, \citenamefont {D'Souza},\ and\ \citenamefont
  {Herrmann}}]{schrenk2012}%
  \BibitemOpen
  \bibfield  {author} {\bibinfo {author} {\bibfnamefont {K.~J.}\ \bibnamefont
  {Schrenk}}, \bibinfo {author} {\bibfnamefont {A.}~\bibnamefont {Felder}},
  \bibinfo {author} {\bibfnamefont {S.}~\bibnamefont {Deflorin}}, \bibinfo
  {author} {\bibfnamefont {N.~A.~M.}\ \bibnamefont {Ara\'ujo}}, \bibinfo
  {author} {\bibfnamefont {R.~M.}\ \bibnamefont {D'Souza}}, \ and\ \bibinfo
  {author} {\bibfnamefont {H.~J.}\ \bibnamefont {Herrmann}},\ }\href@noop {}
  {\bibfield  {journal} {\bibinfo  {journal} {Phys. Rev. E},\ }\textbf
  {\bibinfo {volume} {85}},\ \bibinfo {pages} {031103} (\bibinfo {year}
  {2012})}\BibitemShut {NoStop}%
\bibitem [{\citenamefont {Erd\H{o}s}\ and\ \citenamefont
  {R\'{e}nyi}(1960)}]{ER}%
  \BibitemOpen
  \bibfield  {author} {\bibinfo {author} {\bibfnamefont {P.}~\bibnamefont
  {Erd\H{o}s}}\ and\ \bibinfo {author} {\bibfnamefont {A.}~\bibnamefont
  {R\'{e}nyi}},\ }\href@noop {} {\bibfield  {journal} {\bibinfo  {journal}
  {Publ. Math. Inst. Hungar. Acad. Sci},\ }\textbf {\bibinfo {volume} {5}},\
  \bibinfo {pages} {17} (\bibinfo {year} {1960})}\BibitemShut {NoStop}%
\bibitem [{\citenamefont {Bollob{\'a}s}(1984)}]{bollobas84}%
  \BibitemOpen
  \bibfield  {author} {\bibinfo {author} {\bibfnamefont {B.}~\bibnamefont
  {Bollob{\'a}s}},\ }\href@noop {} {\bibfield  {journal} {\bibinfo  {journal}
  {Trans. Amer. Math. Soc.},\ }\textbf {\bibinfo {volume} {286}},\ \bibinfo
  {pages} {257} (\bibinfo {year} {1984})}\BibitemShut {NoStop}%
\end{thebibliography}%

\end{document}